\documentclass[aps,letter,prx,reprint,superscriptaddress,notitlepage,floatfix,nolongbibliography]{revtex4-2}

\usepackage{graphicx}
\usepackage{multirow}
\usepackage{amsmath,amssymb,amsfonts}
\usepackage{amsthm}
\usepackage{mathrsfs}
\usepackage{xcolor} 
\usepackage{textcomp}
\usepackage{booktabs}

\usepackage{physics}
\usepackage{tikz}
\usepackage{pgfplots}
\usepackage{overpic}
\usepackage{xr}
\usepackage{hyperref}  

\pgfplotsset{compat=1.18}
\usetikzlibrary{calc,positioning,arrows.meta}

\theoremstyle{plain}

\theoremstyle{remark}

\definecolor{NatBlue}{HTML}{0072B2}
\definecolor{NatOrange}{HTML}{E69F00}
\definecolor{NatVermillion}{HTML}{D55E00}
\definecolor{NatGreen}{HTML}{009E73}
\definecolor{NatPurple}{HTML}{CC79A7}
\definecolor{NatGrey}{HTML}{999999}

\begin{document}

\title{%
  Discovering and decoding latent mean-field structure with
  variational autoencoders%
}

\author{Marco Biroli}
\affiliation{%
  Department of Physics and the James Franck Institute,
  University of Chicago, Chicago, IL, USA%
}

\author{Max Welling}
\affiliation{CuspAI, Cambridge, England}
\affiliation{%
  AMLab, University of Amsterdam, Amsterdam, Netherlands%
}

\author{Vincenzo Vitelli}
\affiliation{%
  Department of Physics and the James Franck Institute,
  University of Chicago, Chicago, IL, USA%
}
\affiliation{
Leinweber Institute for Theoretical Physics 
}

\begin{abstract}
Generative models are increasingly used to capture correlations in many-body systems, but the representations they learn remain largely opaque to physical interpretation. Here, we establish an intuitive  criterion that quantifies the capacity of a variational autoencoder (VAE) to faithfully reconstruct the joint probability distribution of a many body system. In a nutshell, a bound on the VAE capacity is obtained by comparing the rate of the latent channel to the bipartite mutual information of the data. Using this bound, we show that the conditionally independent decoder of any successful VAE is structurally identical to a finite-size mean-field factorization. Hence, a successful reconstruction is direct evidence for a latent mean-field theory and the microscopic parameters of that theory can be read off the trained decoder. We validate these conclusions on a hierarchy of solvable models with scalar (Curie-Weiss), vector (Hopfield) and tensor (Maier-Saupe) order parameters, recovering the full Hopfield pattern matrix from equilibrium samples alone. We find that, when applied to Salamander retinal recordings, a two-latent VAE reproduces the population statistics with only two effective collective variables allowing us to recover the `stored patterns' of the neural population and write a generalized Hopfield model which correctly models the experimental data.
\end{abstract}


\maketitle

A long standing challenge in the natural sciences is to predict and understand the behavior of many body systems with complicated correlations among their constituents. Iconic examples range from magnetic materials undergoing phase transitions to information processing in networks of neurons. Collective phenomena in complex systems are often mathematically intractable and require clever approximations and computational methods. The core mathematical difficulty is to estimate the joint probability distribution $p(\vb{x})$  describing a many body system $\vb{x} = \{x_1, x_2, \cdots, x_N\}$. Typically $\vb{x}$ is a collection of $N$ variables with complicated correlations resulting from interactions among microscopic constituents. Consequently, the probability distribution $p(\vb{x})$ is not factorizable,
\begin{equation} \label{eq:not-independent}
    p(\vb{x}) \neq \prod_{i = 1}^N p_i(x_i) \;.
\end{equation}

The rise of artificial intelligence has provided a new set of computational tools to tackle the many body problem. For example, by automating the identification of collective variables or order parameters that reduce the dimensionality of the system under investigation \cite{Mehta2019,Carrasquilla2017,Wetzel2017,Iten2020}. Although very powerful, machine learning tools are often applied to scientific problems in a heuristic fashion without establishing apriori the conditions that these complex phenomena or datasets need to satisfy for the chosen ML approach to work.

\vspace{0.2cm}

\begin{figure*}[t]
    \centering
    \includegraphics[width=0.9\textwidth]{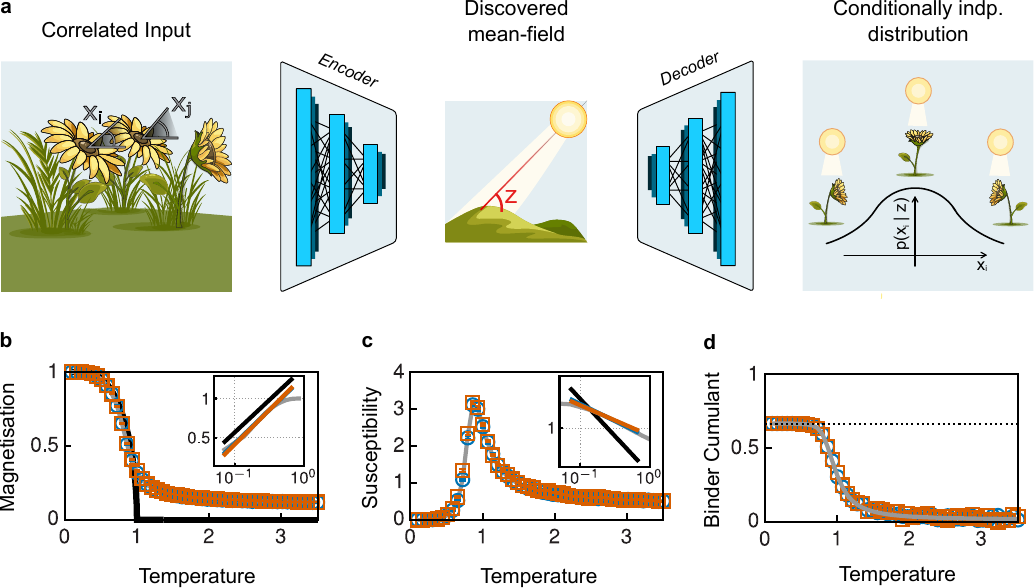}
    \caption{{\bf Discovering mean-field structures.} The top row ({\bf a}) schematically illustrates our approach on a toy example. Suppose we had snapshots of sunflower fields. Training a regular low-dimensional VAE on these images will discover the effective {\it mean-field} acting on these flowers, i.e. the angle of incidence of the sun, despite no picture of the sun every being shown during training. Applying the same method on computer generated equilibrium samples of the Curie-Weiss model of magnetism allows us to recover this mean field model exactly. On the bottom row ({\bf b-c-d}) we show how VAE samples (orange squares) recover the true distribution (blue circles) which matches the theory (gray line) for a variety of observables. Namely, the average magnetization ({\bf b}), the susceptibility ({\bf c}) and the Binder cumulant ({\bf d}). The thermodynamic mean-field prediction is given by a black line. The insets show a log-log zoom of the plot close to the critical temperature.}
    \label{fig:scalar}
\end{figure*}

Energy-based and generative models such as restricted Boltzmann machines \cite{TubianaMonasson2017,MehtaSchwab2014,Decelle2021}, variational autoregressive networks \cite{Wu2019}, diffusion models \cite{Sohl2015,Ho2020} and Variational AutoEncoders (VAEs) \cite{Kingma2014} have been used with great success in several fields including neuroscience \cite{Pandarinath2018}, physics \cite{Wetzel2017,Walker2020,Iso2018,Cossu2019,Cristoforetti2017, CoccoMonasson2011, NguyenBergZecchina2017, MezardMontanari2009}, biophysical modeling \cite{Torlai2016} and computer science \cite{Kingma2014,Goodfellow2016}.
VAEs have limitations and known failure cases \cite{Zhao2017,Larsen2016} where they are unable to correctly approximate $p(\vb{x})$.
In this paper, we will show that VAEs are structurally identical to a finite-size mean-field factorization of the joint distribution, as summarized in Fig.~\ref{fig:scalar}. We will derive when VAEs can succeed in describing correlated systems, or more precisely, when they will necessarily fail irrespective of algorithm implementation or hyperparameter selection. 

\vspace{0.2cm}

The equivalence between VAEs and mean-field models has three consequences. 
First, low-dimensional VAEs provably fail on systems with no mean-field description, such as the two-dimensional Ising model (Sec. \ref{sec:scalar}). 
Second, a successful VAE reconstruction is evidence for a latent mean-field theory, which we demonstrate on the Curie-Weiss (Sec. \ref{sec:scalar}), Hopfield (Sec. \ref{sec:vector}), and Maier-Saupe (Sec. \ref{sec:tensor}) models, spanning scalar, vector, and tensor order parameters. 
Third, the microscopic parameters of the corresponding mean-field theory can be recovered directly from the trained decoder. We illustrate this by recovering Hopfield patterns from equilibrium spin samples. 
We conclude by applying these ideas to retinal population recordings under naturalistic stimulation \cite{Schneidman2006}. The trained VAE uncovers two low-dimensional mean-field codes organizing the $40$ ganglion cells, leading to a generalized Hopfield model \cite{krotov2016dense} with fitted parameters which accurately reproduce the experimental data.

\section{Conditional independence: the key VAE assumption}
\label{sec:vae}

\begin{figure*}[t]
    \centering
    \includegraphics{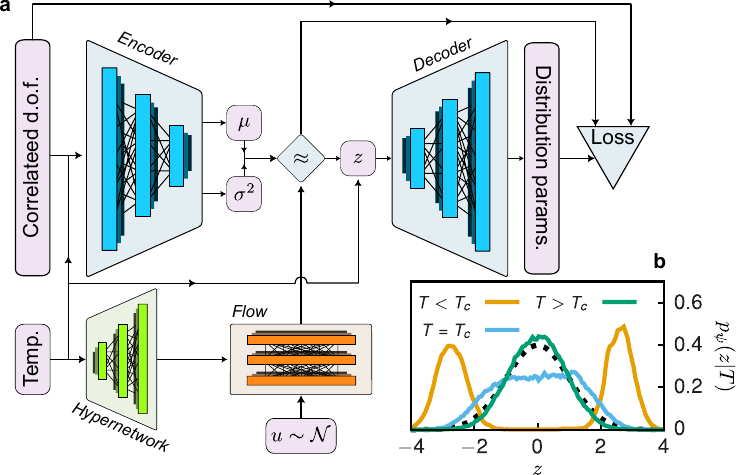}
    \caption{{\bf VAE architecture.} A detailed view of the conditional VAE architecture ({\bf a}). The correlated degrees of freedom (e.g. spins, neurons, nematics) are concatenated with the temperature at which they were sampled and given to an encoder (blue) which outputs the mean and variance of the posterior distribution $q(\vb{z} | \vb{x}) \sim \mathcal{N}(\mu, \sigma^2)$. The temperature is given to a hyper-network (green) which predicts the weights of a spline flow network (orange) which builds the prior distribution $p_\psi(z)$ through invertible transformations of a Gaussian random variable $u \sim \mathcal{N}(0, 1)$. With the reparameterization trick \cite{Kingma2014} ($\approx$) the latent $z$ is sampled from the encoder prediction and given to the decoder (blue) which predicts the parameters of the conditionally independent distributions $p_i(x_i | \vb{z})$. The loss is computed from the likelihood of the original input given the decoded distribution and the Kullback-Leibler divergence between the prior and posterior distribution. On the bottom right ({\bf b}), we show the prior $p_\psi(z)$ which was learned for the Curie-Weiss model. As expected above the critical temperature $T > T_c$ the distribution is uni-modal and approximately Gaussian (black dashed curve), it flattens out as $T \approx T_c$ and becomes clearly bi-modal for $T \ll T_c$.}
    \label{fig:vae_arch}
\end{figure*}

For a complete introduction to VAEs please refer to Ref.~\cite{VAE_book}, we will only sparsely introduce them here in layman's terms. Variational AutoEncoders (depicted in Fig.~\ref{fig:scalar} and shown in detail in Fig.~\ref{fig:vae_arch}) are generative machine learning models built out of three main components: an encoder, a prior and a decoder. The encoder maps data $\vb{x}$ from its original input space into some latent representation $\vb{z}$. Inversely, the decoder maps latent space samples $\vb{z}$ back to data $\vb{x}'$. Finally, the prior defines the geometry of the latent space.

\vspace{0.2cm}

The intuition behind VAEs is that latent variables $\vb{z}$ can absorb most of the correlations in $\vb{x}$ into a more regular space (see Fig.~\ref{fig:scalar} for a toy example). For example, imagine $\vb{z}$ as a variable encoding the pose, lighting, or some facial features of a portrait in such a way that $\vb{z}$ fully characterizes the portrait. Then the individual pixels $\vb{x}$ of the image can be fully recovered, up to independent fluctuations, from $\vb{z}$. Mathematically, we decompose the joint probability distribution function (jpdf) as
\begin{equation} \label{eq:total-prob}
    p(\vb{x}) = \int \dd \vb{z} \; p(\vb{z}) p(\vb{x} | \vb{z}) \;,
\end{equation}
where $p(\vb{z})$ is the jpdf of the latent variables (often assumed to be Gaussian and isotropic) and $p(\vb{x} | \vb{z})$ is the conditional probability of observing $\vb{x}$ given $\vb{z}$.

\vspace{0.2cm}

In practice, a VAE (shown in Fig.~\ref{fig:vae_arch}) works as follows. First, an encoder network $\phi$ outputs the parameters of a Gaussian posterior $q_\phi(\vb{z}|\vb{x}) = \mathcal{N}(\vb{\mu}_\phi(\vb{x}), \vb{\sigma}_\phi(\vb{x}) \vb{I})$ which, when averaged over the data distribution $p(\vb{x})$, attempts to approximate a learned prior distribution $p_\psi(\vb{z})$ generated from a conditional spline flow network \cite{Rezende2015,Durkan2019}. Second, a decoder network $\theta$ attempts to approximate the conditional distribution $p(\vb{x} | \vb{z})$. The three networks $(\phi, \psi, \theta)$ are jointly trained by maximizing the evidence lower bound (ELBO) on the marginal log-likelihood,
\begin{align} 
    \log p_\theta(\vb{x}) \;\geq\; \langle \log &p_\theta(\vb{x}|\vb{z}) \rangle_{q_\phi(\vb{z}|\vb{x})} \nonumber\\
    &- D_\text{KL}\!\left(q_\phi(\vb{z}|\vb{x}) \,\|\, p_\psi(\vb{z})\right) \;,\label{eq:elbo}
\end{align}
where $D_{\rm KL}(q || p)$ is the Kullback-Leibler divergence defined as
\begin{equation} \label{eq:def-KL}
    D_{\rm KL}(P || Q) = \int \dd \vb{x} \; P(\vb{x}) \log \frac{P(\vb{x})}{Q(\vb{x})} \;.
\end{equation}
The bound in Eq.~(\ref{eq:elbo}) is tight when $q_\phi(\vb{z}|\vb{x})$ matches the true posterior $p_\theta(\vb{z}|\vb{x})$ \cite{Kingma2014}. 

\vspace{0.2cm}

Note that Eq.~\eqref{eq:total-prob} is the law of total probability and holds for any joint $p(\vb{x})$. For $p(\vb{z})$ to absorb the leading correlations in $p(\vb{x})$, we need $p(\vb{x} \mid \vb{z})$ to be of a simpler form than $p(\vb{x})$ and easier to learn. The most common simplification is to assume that
\begin{equation} \label{eq:cond-indp}
    p_\theta(\vb{x} | \vb{z}) = \prod_{i = 1}^N p_\theta^{(i)}(x_i | \vb{z}) \;.
\end{equation}

Mathematically, the variables $\vb{x}$ are said to be {\it conditionally independent} \cite{biroli2025strongly} given the latent variables $\vb{z}$. Substituting Eq.~(\ref{eq:cond-indp}) into Eq.~(\ref{eq:total-prob}) yields the c.i.i.d. ansatz, 
\begin{equation} \label{eq:jpdf-cond-indp}
    p_\theta(\vb{x}) = \int \dd \vb{z} \; p_\psi(\vb{z}) \prod_{i = 1}^N p_\theta^{(i)}(x_i | \vb{z}) \;.
\end{equation}
The seemingly inconspicuous steps that were taken between Eq.~(\ref{eq:total-prob}) and Eq.~(\ref{eq:jpdf-cond-indp}) are crucial. Although any distribution $p(\vb{x})$ can always be written as in Eq.~\eqref{eq:total-prob} only a select few can be written as in Eq.~\eqref{eq:jpdf-cond-indp} with non-degenerate distributions and a low-dimensional $\vb{z}$. The conditional independence assumption which we make in Eq.~\eqref{eq:jpdf-cond-indp} is a strong structural constraint on the behavior of the variables $\vb{x}$. We now show that this conditional-independence assumption is structurally identical to a finite-size mean-field factorization (Sec.~\ref{sec:equivalence}).

\begin{figure*}[tp] 
    \centering
    \includegraphics{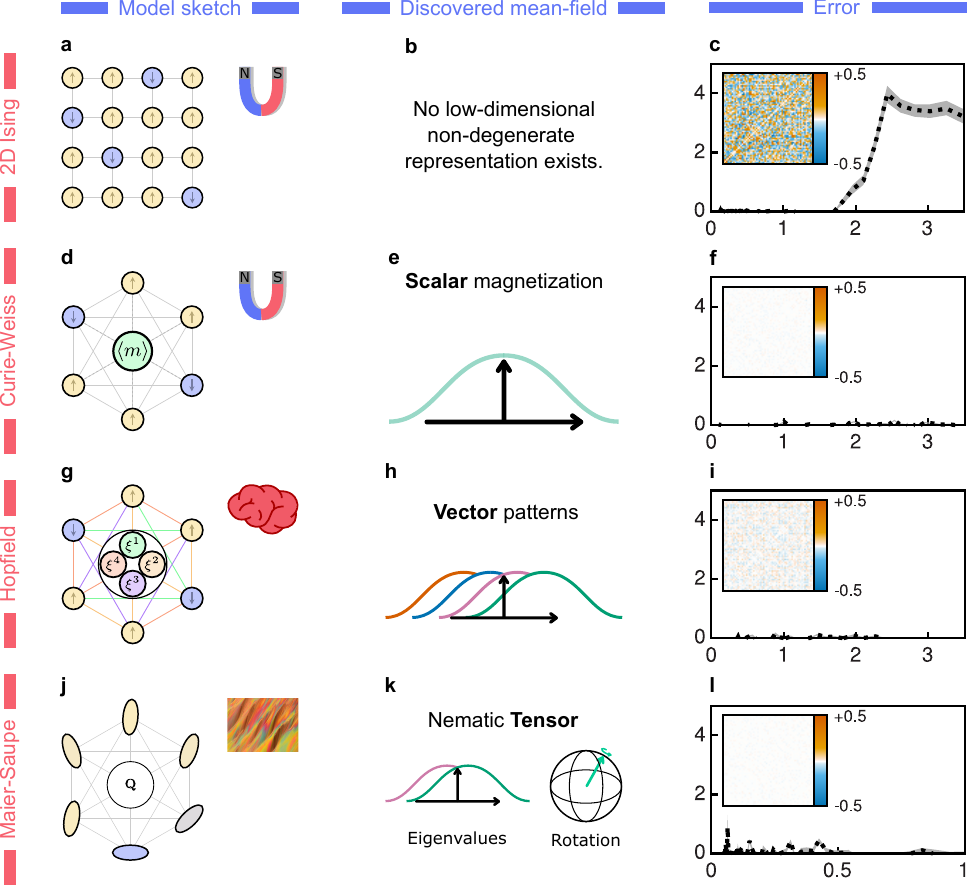}
    \caption{{\bf Hierarchy of models.} Each row is one system; from top to bottom: the two-dimensional Ising model ({\bf a-b-c}), the Curie-Weiss model ({\bf d-e-f}), the Hopfield model with $P = 4$ stored patterns ({\bf g-h-i}), and the Maier-Saupe nematic ({\bf j-k-l}). The left column ({\bf a-d-g-j}) shows a sketch of the model under investigation. All degrees-of-freedom are correlated (exemplified by the gray lines) and hence not independent. The middle column ({\bf b-e-h-k}) shows the uncovered mean-field description that the VAE extracts from the correlated data. The right column ({\bf c-f-i-l}) displays the conditional total correlation ${\rm TC}_{|z}$ of Eq.~(\ref{eq:TC_KL}), estimated through the variational lower bound of Eq.~(\ref{eq:f-divergence-loss}) as a function of temperature $T$. The shaded band gives the bootstrap standard deviation. The inset in each row shows the contribution of every two-point observable to the mismatch between the true data distribution and the factorized VAE approximation. For the two-dimensional Ising model (top row) the residual grows and seems to peak around $\sim T_c \approx 2.27$, and the inset matrix displays the structured two-point correlations that a conditionally independent decoder fails to capture: this is the canonical illustration of the violation of Eq.~(\ref{eq:capacity}). For every other model the conditional total correlation is indistinguishable from zero.}
    \label{fig:hierarchy}
\end{figure*}

\section{Equivalence with finite-size mean-field theory}
\label{sec:equivalence}

{\bf Mean-field approximation. } The mean-field approximation is a powerful tool in statistical physics used to approximate $p(\vb{x})$ when it is intractable. Mean-field theory is an umbrella term loosely used for any theory which absorbs all
correlations into just a few degrees of freedom \cite{sethna,Kadanoff2000}. To clarify what the mean-field approximation entails we are going to reintroduce it precisely for the Ising model. The Ising model is the simplest statistical model of magnetism which represents a magnet as a $d$-dimensional cubic lattice where each vertex $i$ is a microscopic spin whose up/down state is represented by $x_i \in \{1, -1\}$. The total energy of the magnet is given by
\begin{equation} \label{eq:H-ising}
    H = - J \sum_{i \; {\rm n.n.} \; j} x_i x_j \;,
\end{equation}
where the sum runs over all nearest neighbors (n.n) pairs. Statistical mechanics \cite{sethna, Kadanoff2000} tells us that at equilibrium the jpdf is given by the Gibbs-Boltzmann distribution
\begin{equation} \label{eq:gibbs}
    p(\vb{x}) = \frac{1}{Z} e^{-\beta H} \;,
\end{equation}
where $\beta = 1/T$ is the inverse temperature and 
\begin{equation} \label{eq:partition}
    Z = \int \dd \vb{x} \; e^{-\beta H} \;,
\end{equation}
is the partition function. In one and two dimensions an exact solution for $Z$ is known \cite{Onsager1944}, whereas for three-dimensional lattices no closed form is known and $Z$ must be evaluated numerically or approximately \cite{Kadanoff2000} and hence we need to resort to approximations. Let us focus on an arbitrary spin $i$ and decompose the Hamiltonian as
\begin{equation} \label{eq:H-decomposition}
    H = -J  x_i\sum_{j {\rm \; n.n. \; } i} x_j + {\rm const.} \equiv H_i(h_i) + {\rm const} \;,
\end{equation}
i.e. each of the $x_i$ are random variables subject to a random local field $h_i = \sum_{j {\rm \; n.n. \;} i} x_j$ resulting from the interaction with each of their nearest neighbors. Since $i$ is chosen arbitrarily we can further assume that there is no specificity to $i$ and that the local field $h_i$ is everywhere well described by a single mean field $h$. This is the mean field assumption. However, oftentimes we further assume that $i$ has enough neighbors and that the mean-field $h$ is self-averaging. This would allow us to treat $h$ deterministically with $h \longrightarrow \langle h \rangle$ in which case
\begin{equation}\label{eq:indp-mean-field}
    p_{\rm VI}(\vb{x}) = \prod_{i = 1}^N \frac{1}{Z_i}e^{-\beta H_i(\langle h \rangle)} \;.
\end{equation}
We can immediately see that Eq.~\eqref{eq:indp-mean-field} is often a very crude approximation of Eq.~\eqref{eq:gibbs} since it assumes independence and neglects {\rm all correlations} $\langle x_i x_j\rangle - \langle x_i \rangle\langle x_j \rangle = 0$. In the context of variational inference, mean-field often refers to the much less expressive Eq.~\eqref{eq:indp-mean-field}. In some cases the approximation is justified. However, for numerous realistic systems, e.g. finite-size ones, either the thermodynamic hypothesis or the self-averaging hypothesis or both are violated. However, we can still make the mean-field assumption without taking the extra step of assuming $h \to \langle h \rangle$ in which case 
\begin{equation} \label{eq:CI-mean-field}
    p_{\rm MF}(\vb{x}) = \int \dd h \; p(h) \prod_{i = 1}^N \frac{1}{Z_i}e^{-\beta H_i(h)} \;.
\end{equation}
We can now see that Eq.~\eqref{eq:CI-mean-field} is a special instance of  Eq.~\eqref{eq:jpdf-cond-indp}, which is much more expressive than Eq.~\eqref{eq:indp-mean-field}. Most importantly, notice that using Eq.~\eqref{eq:CI-mean-field} we can describe non-zero correlations $\langle x_i x_j \rangle - \langle x_i \rangle\langle x_j\rangle \neq 0$.

\vspace{0.2cm}

To illustrate the difference between Eq.~\eqref{eq:indp-mean-field} and Eq.~\eqref{eq:CI-mean-field} we take the Curie-Weiss model which is the prototypical mean-field model defined as
\begin{equation} \label{eq:H-CW}
    H_{\rm CW} = -\frac{J}{2N} \sum_{i \neq j} x_i x_j = -\frac{J}{2N} \left[ \left(\sum_{i = 1}^N x_i\right)^2 - N\right] \;.
\end{equation}
For this particular Hamiltonian we can re-write the jpdf as 
\begin{equation} \label{eq:jpdf-CW}
    p_{\rm CW}(\vb{x}) = \frac{1}{Z} e^{\frac{\beta J}{2 N} \left(\sum_{i} x_i\right)^2} \;.
\end{equation}
Using a Gaussian integral identity we can get rid of the square in the exponential
\begin{equation}
    p_{\rm CW}(\vb{x}) = \int \frac{\dd z}{Z \sqrt{\pi}} e^{-\beta J N z^2 / 2} \prod_{i = 1}^N e^{\beta J x_i z} \;,
\end{equation}
which can immediately be re-written as
\begin{equation} \label{eq:jpdf-CW-CI}
    p_{\rm CW}(\vb{x}) = \int \dd z \; p(z) \prod_{i = 1}^N p(x_i | z) \;,
\end{equation}
where 
\begin{equation} \label{eq:pxz-CW}
    p(x_i | z) = \frac{e^{\beta J x_i z}}{2 \cosh(\beta J z)} \;,
\end{equation}
and 
\begin{equation} \label{eq:pz-CW}
    p(z) = \frac{2^N \cosh(\beta J z)^N}{Z \sqrt{\pi}} e^{-\beta J N  z^2/2} \;.
\end{equation}

Note that the mean field variable $z$ is in general stochastic. Only when $N \to +\infty$ can we perform the saddle-point approximation of $p(z)$ whose extrema are given by the famous identity
\begin{equation}
    z = \tanh(\beta J z) \;,
\end{equation} 
to recover Eq.~\eqref{eq:indp-mean-field}. Hence, even in the Curie-Weiss model, we see that Eq.~\eqref{eq:CI-mean-field} has the correct structure. The usual form given in Eq.~\eqref{eq:indp-mean-field} is only valid when taking the $N \to +\infty$ limit on top of the mean-field approximation. Notice that the critical temperature $\beta_c = 1$ of the Curie-Weiss model also corresponds to the temperature at which $p(z)$ goes from being unimodal to becoming bimodal, as can be seen in Fig.~\ref{fig:vae_arch}.

\section{A measure of success}

We now aim to provide a measurable estimator that allows us to verify whether the latent-variable description has correctly captured the correlations of the many-body system. The relevant quantity for us to study is the total correlation
\begin{equation} \label{eq:TC_KL}
    {\rm TC}_{|\vb{z}} = D_{\rm KL}\left[ p(x_1, \cdots, x_N | \vb{z}) \; \Big|\Big| \; \prod_{i = 1}^N p(x_i | \vb{z}) \right] \;,
\end{equation}
The total correlation measures `how far' the joint probability density function is from its independent counterpart. Equivalently, it can also be defined in terms of the information entropy $S(x_i | \vb{z})$ as
\begin{equation} \label{eq:TC_S}
    {\rm TC}_{|\vb{z}} = \sum_{i = 1}^N S(x_i | \vb{z}) - S(x_1, \cdots, x_N | \vb{z}) \;.
\end{equation}
The physical interpretation is that ${\rm TC}_{|\vb{z}}$ measures the amount of information contained in the correlations of the system at hand. In other words, how much information is the independent model failing to capture. In our case, since we expect that $\vb{z}$ contains all of the relevant correlation information of the system we expect that ${\rm TC}_{|\vb{z}} \simeq 0$. Any deviation from 0 would correspond to the amount of information that the VAE latent $\vb{z}$ has failed to capture. Modern techniques \cite{Xu2025} allow us to bound ${\rm TC}_{|\vb{z}}$ accurately by using a neural network to learn the relative density ratio 
\begin{equation} \label{eq:RDR}
    r(x_1, \cdots, x_N | \vb{z}) = \frac{p({\vb x} | \vb{z})}{\frac{1}{2} \left[ p({\vb x} | \vb{z}) + \prod_{i = 1}^N p(x_i | \vb{z}) \right]} \;,
\end{equation}
see the Methods (Sec.~\ref{sec:meth:TC}) for details on this procedure. Learning the relative density ratio in Eq.~(\ref{eq:RDR}) has an added bonus compared to simply reporting integrated metrics such as the total correlation in Eq.~(\ref{eq:TC_KL}). 
Since $r(\vb{x} | \vb{z})$ is defined for any $\vb{x}$, we can fit a simple ridge regression to see what observables of $\vb{x}$ contribute most to the mismatch between $p(\vb{x} | \vb{z})$ and $\prod_{i = 1}^N p_i(x_i | \vb{z})$.
We can measure exactly {\it which observables the VAE fails to capture}. In our case, we will measure the contribution of all one-point functions $\langle x_i \rangle$ and two-point functions $\langle x_i x_j \rangle$ by regressing $x_i - \langle x_i \rangle$ and $(x_i - \langle x_i \rangle)(x_j - \langle x_j \rangle)$ respectively which are shown in the insets of Fig.~\ref{fig:hierarchy}(c-f-i-l). This is equivalent to looking at the residuals of a linear regression instead of simply looking at an integrated metric like $R^2$. 

\section{When can a VAE fit
\texorpdfstring{$p(\vb{x})$}{p(x)}?}
\label{sec:capacity}

The conditional independence assumption of Eq.~(\ref{eq:jpdf-cond-indp}) enforces all message-passing between $x_i$ and $x_j$ to be routed through $\vb{z}$ for any $i$ and $j$. If we cut the entire system $\{1, \cdots, N\}$ into two disjoint pieces $A \sqcup B = \{1, \cdots, N\}$ then all interactions between $\vb{x}_A = \{x_i : i \in A\}$ and $\vb{x}_B$ must be mediated by $\vb{z}$. Consequently, all the information shared between $\vb{x}_A$ and $\vb{x}_B$ must be contained in the joint description of $\vb{x}$ and $\vb{z}$. Mathematically, this yields the inequality
\begin{equation} \label{eq:markov-chain}
    I(\vb{x}, \vb{z}) \geq I(\vb{x}_A, \vb{x}_B) \;,
\end{equation}
where $I(x, y) = H(x) - H(x | y)$ is the mutual information of $x$ and $y$. This inequality holds for any $A$ and $B$, we can tighten the bound by maximizing 
\begin{equation} \label{eq:bip}
    I(\vb{x}, \vb{z}) \geq I_{\rm bip}(p) \equiv \max_{A \sqcup B} I(\vb{x}_A, \vb{x}_B) \;.
\end{equation}
The quantity $I_{\rm bip}(p)$ is a physical property of the system we study which measures {\it the amount of information required to describe the correlations of the physical system.}

\vspace{0.2cm}

On the other hand, a famous identity, called ELBO surgery, from the VAE literature \cite{HoffmanJohnson2016,Alemi2018} states 
\begin{equation} \label{eq:elbo-surgery}
    R = I(\vb{x}, \vb{z}) + D_{\rm KL}(q_\phi(\vb{z}) || p_\psi(\vb{z})) \;,
\end{equation}
where 
\begin{equation} \label{eq:rate-kl}
    R \equiv \langle D_{\rm KL}(q_\phi(\vb{z} | \vb{x}) || p_\psi(\vb{z}) \rangle_{p(\vb{x})} \;.
\end{equation}
This equation can be understood as follows. The prior $p_\psi(\vb{z})$ can be interpreted as the quantity defining the `geometry' of the latent space. On the other hand, $q_\phi(\vb{z})$ is the encoder's approximation of this geometry. 
The rate $R$ measures how `uniquely' each data point $\vb{x}$ is encoded in the true latent space $p_\psi(\vb{z})$. Uniqueness can come in two forms. Either $\vb{z}$ contains enough information about $\vb{x}$ that two different samples $\vb{x}, \vb{x}'$ can be discriminated into $\vb{z}, \vb{z}'$, this is measured by $I(\vb{x}, \vb{z})$. Or the approximate latent space $q_\phi(\vb{z})$ does not match the true latent space $p_\psi(\vb{z})$ making encoded samples trivially `unique', since they will not fit in the true geometry of the latent space. If the VAE encoder is trained properly, this second term should vanish and the approximate geometry $q_\phi(\vb{z})$ should match the true geometry $p_\psi(\vb{z})$ \cite{HoffmanJohnson2016,Alemi2018,Tishby2015}. Then $R$ can be interpreted as {\it the amount of information that the trained VAE encoder is able to pack into the latent space.}

\vspace{0.2cm}

Matching Eq.~(\ref{eq:bip}) and Eq.~(\ref{eq:elbo-surgery}) leads to a criterion necessary for a VAE to be able to successfully model a system
\begin{equation} \label{eq:capacity}
    R \gtrsim I_{\rm bip}(p) \;.
\end{equation}
In other words, the amount of information $R$ that the VAE is able to encode into the latent space has to be greater than the amount of information $I_{\rm bip}(p)$ required to describe the correlations of $p(\vb{x})$. For a large class of VAE encoders the rate $R$ can be approximated by 
\begin{equation}
    R \simeq d \log (1 / \sigma) \;,
\end{equation}
where $d$ is the dimension of the latent space and $\sigma$ is the average precision of the VAE encoder. A more formal and precise description of this argument is provided in the Methods.

\vspace{0.2cm}

Combining the mean-field description given in Sec.~\ref{sec:equivalence} with Eq.~(\ref{eq:capacity}) we can now write formally the three main conclusions put forward in this paper.
\begin{enumerate}
  \item[{\bf C1.}] If the joint distribution $p(\vb{x})$ admits no mean-field description (i.e. its correlations cannot be captured by a couple of low-dimensional order parameters), then $I_{\rm bip}(p) \sim \mathcal{O}(N^\alpha)$ and VAEs with a factorizable decoder, low-dimensional latents ($d \sim \mathcal{O}(1)$) and a smooth regular prior ($\sigma \neq 0$) will fail to reconstruct $p(x)$.
  \item[{\bf C2.}] If a VAE with a factorizable decoder, low-dimensional latents and smooth regular prior successfully reconstructs $p(\vb{x})$, then $p(\vb{x})$ must admit a non-trivial mean-field description.
  \item[{\bf C3.}] If the VAE has been trained successfully, the microscopic mean-field parameters can be recovered from the trained network.
\end{enumerate}
The remainder of the paper instantiates each of these three corollaries on a hierarchy of synthetic solvable models (Sec.~\ref{sec:validation}) and on experimental retinal recordings (Sec.~\ref{sec:retina}).

\begin{figure*}[tp]
    \centering
    \includegraphics[width=\textwidth]{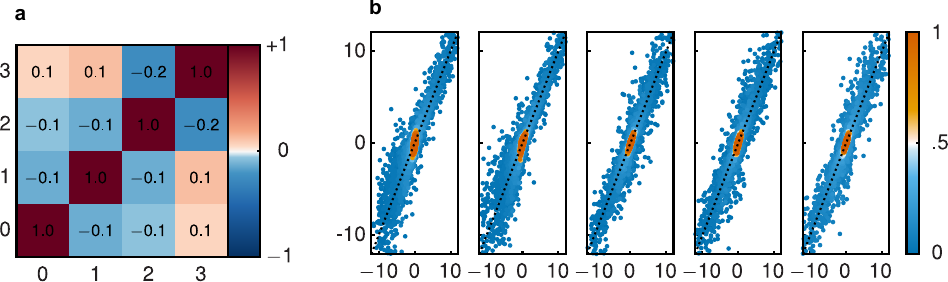}
    \caption{{\bf Recovery of microscopic mean-field parameters from the trained decoder.} On the left ({\bf a}), Hopfield pattern overlap matrix. Each entry $(\mu, \nu)$ is the per-site overlap $\langle \hat{\vb{\xi}}^\nu | \vb{\xi}^\mu \rangle / N$ between the $\mu$-th true pattern and the $\nu$-th pattern extracted from the VAE. The matrix is diagonal with unit modulus along the diagonal proving that all $P = 4$ patterns are perfectly recovered from equilibrium samples alone. On the right ({\bf b}), each panel is a scatter of one independent component $Z_{\alpha\beta}$ (from left to right $Z_{00}$, $Z_{11}$, $Z_{01}$, $Z_{02}$, $Z_{12}$) of the nematic tensor $\vb{Z}$ against the MLP prediction from the encoder posterior mean $\vb{\mu}(\hat{\vb{x}})$; the dotted line is $\vb{Z} = \hat{\vb{Z}}$ and points are colored by the temperature $T/J$ ranging from $0.05$ to $1$ (color bar on the right). The collapse onto the diagonal in every panel, with per-component coefficient of determination $R^2 \geq 0.90$, shows that the $5$-dimensional latent is directly related to the physical nematic tensor.}
    \label{fig:vector-tensor}
\end{figure*}

\section{Validation: a hierarchy of order parameters}
\label{sec:validation}

We now check Eq.~(\ref{eq:capacity}) on three solvable systems whose order parameters are respectively scalar (\ref{sec:scalar}), vectors (\ref{sec:vector}) and tensors (\ref{sec:tensor}) and on a system which does not admit a mean-field description (\ref{sec:scalar}). Fig.~\ref{fig:hierarchy} summarizes the outcome. For each use case, the VAE is instantiated with a CNN encoder/decoder and latent dimensions matching the expected dimensions from theory, i.e. $1, P$ and $5$ respectively for Curie-Weiss, Hopfield and Maier-Saupe. The decoders output logits for the spin-systems and the parameters of a Bingham distribution for Maier-Saupe. See the Methods~\ref{sec:meth:VAE} for more implementation details.

\subsection{Scalar order parameter: Ising versus Curie-Weiss}
\label{sec:scalar}

We first illustrate the equivalence on the two-dimensional Ising and Curie-Weiss models, which have also been studied with VAEs and related generative architectures in earlier work \cite{Wetzel2017,Walker2020,Iso2018}. However, their analysis is usually restricted to one-point observables (magnetization / susceptibility / binder cumulant). This hides the true failure of the VAEs' reconstruction of 2D Ising which is contained in two-point functions. We accurately quantify this failure mode using ${\rm TC}_{|\vb{z}}$. In two dimensions the Ising model has no mean-field description. Mean-field becomes accurate only in the infinite dimensional limit. On the other hand, the Curie-Weiss model is always mean-field by construction but the fully independent factorization in Eq.~\eqref{eq:indp-mean-field} becomes valid only in the thermodynamic limit. We generated $10^{5}$ independent two-dimensional Ising equilibrium samples (resp. Curie-Weiss equilibrium samples) for $50$ different temperatures ranging from $0.1$ to $3.5$ (with $J = 1$) using the Wolff cluster algorithm \cite{Wolff1989} (resp. using direct sampling). We then trained a VAE with a latent dimension of $1$ to reconstruct the equilibrium samples. We see in Fig.~\ref{fig:scalar}(b,c,d) that the VAE reproduces $\langle |m| \rangle(T)$ for CW across the entire transition region. The agreement extends to higher moments: at $T \ll T_c$ the Binder cumulant matches the universal value $U_4^\star \approx 0.66$, and the susceptibility is also perfectly reconstructed. The VAE is able to perfectly reconstruct all single-point observables, but it is not able to reconstruct two-point observables for 2D Ising (such as the correlation length $C(r) = \langle x(0) x(r) \rangle$), as can be seen in the top row of Fig.~\ref{fig:hierarchy}. The conditional total correlation for 2D Ising increases with temperature and seems to peak around $T_c$. In contrast, the Curie-Weiss model is perfectly captured by the VAE at all temperatures.

\subsection{Vector order parameter: the Hopfield model}
\label{sec:vector}

The Hopfield model extends the Curie-Weiss framework to the storage of $P$ binary patterns $\vb{\xi}^\mu \in \{\pm 1\}^N$ through Hebbian couplings \cite{Hopfield1982}
\begin{equation}\label{eq:H-hopfield}
    H = -\frac{J}{2N}\sum_{i \neq j}\sum_{\mu=1}^{P} \xi_i^\mu \xi_j^\mu\, x_i x_j \;.
\end{equation}
The same Hubbard-Stratonovich transformation that produced Eq.~\eqref{eq:jpdf-CW-CI} for the Curie-Weiss model generalizes here, introducing a $P$-dimensional latent field $\vb{z}$ and yielding the conditional
\begin{equation}\label{eq:pxz-hopfield}
    p(x_i \mid \vb{z}) = \sigma\!\left(\beta x_i J\,(\vb{\xi} \vb{z})_i\right) \;, \qquad (\vb{\xi} \vb{z})_i \equiv \sum_{\mu = 1}^{P} \xi_i^\mu z_\mu \;,
\end{equation}
where $\sigma(u) = e^{u}/({2 \cosh(u)})$. Curie-Weiss is can be recovered by setting $P = 1$. Arbitrary $P$ remains in the same universality class as Curie-Weiss up to the Amit-Gutfreund-Sompolinsky capacity bound $\alpha = P/N \lesssim 0.138$ \cite{AmitGutfreundSompolinsky1985}.

\vspace{0.2cm}

We trained a VAE with latent dimension $P = 4$ on $N = 64$ equilibrium samples produced by a Hubbard-Stratonovich Gibbs sampler (see Methods~\ref{sec:meth:sampling}) at temperatures spanning the retrieval transition. The VAE reproduces the mean-field prediction for the maximum pattern overlap $\langle |m|_{\max}\rangle(T)$ across the entire retrieval transition, including both the condensed low-temperature plateau and the $1/\sqrt{N}$ isotropic baseline above $T_c/J = 1$ (C2). The susceptibility and the Binder cumulant are also perfectly recovered by the VAE and the conditional total correlation is identically $0$, demonstrating that the VAE has fully captured all the correlations in the data.

\vspace{0.2cm}

Furthermore, we can use the trained VAE to solve the inverse-Ising problem \cite{CoccoMonasson2011,NguyenBergZecchina2017}. The key observation is that if the VAE trained successfully, the decoder's logit-space Jacobian
\begin{equation}\label{eq:hopfield-jacobian}
    \chi_{n\mu} \equiv \pdv{f_\theta(\vb{z})_n}{z_\mu} \;, \quad f_\theta(\vb{z})_n \equiv \log\frac{p_\theta(x_n = +1 \mid \vb{z})}{p_\theta(x_n = -1 \mid \vb{z})} \;,
\end{equation}
must equal $2\beta J\,(\vb{\xi} R^{-1})_{n\mu}$ for some unknown rotation $R$, reflecting the freedom to reparametrize $\vb{z}$. Although the inner-computations of the decoder can be a black-box via automatic differentiation we can always access the jacobian $\chi_{n\mu}$. Since the patterns $\vb{\xi}$ are binary $\{+1, -1\}$ vectors, we can use this strong constraint to eliminate $R$ and recover them exactly from $\vb{\chi}$ \cite{Hyvarinen1999}. From equilibrium samples alone, the VAE therefore reconstructs the \emph{unknown} microscopic couplings $\vb{\xi}$ (Fig.~\ref{fig:vector-tensor}a), demonstrating C3. Experimentally, the VAE is able to recover with $100\%$ accuracy at any temperature all $P$ patterns for $P = 4$ and $N = 64$. In fact, the VAE is able to recover with good accuracy ($>95\%$) patterns way beyond the Amit-Gutfreund-Sompolinsky bound, we start seeing some degradation only at $P = 16$, i.e. $\alpha = 0.25$.

\begin{figure*}[tp]
    \centering
    \includegraphics[width=\textwidth]{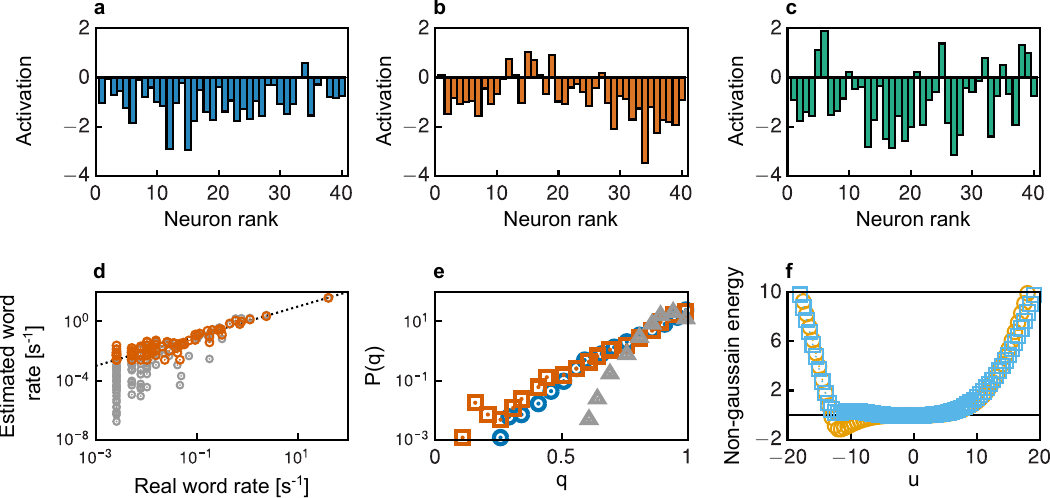}
    \caption{{\bf Application to experimental Salamander retinal recordings.} On the top row ({\bf a-b-c}), we show the stored Hopfield patterns ({\bf a-b}) and external field ({\bf c}) recovered from the trained VAE decoder weights. We can see a large fraction of mostly `silent' neurons, strongly inhibited by the external field and two non-trivial patterns. On the bottom row ({\bf d-e-f}), we show a series of observables for the neural dataset. On the left ({\bf d}), is the per-word firing rate on the salamander retina dataset of Ref.~\cite{Schneidman2006} for the trained VAE (orange) and an independent model baseline (gray). The VAE points scatter tightly around the diagonal across six decades while the independent model systematically undershoots high-rate words. In the middle ({\bf e}), we show the overlap distribution $P(q)$ of Eq.~(\ref{eq:overlap}) between pairs of population codewords from $(i)$ a Monte-Carlo bootstrap of the held-out test set (blue circles), $(ii)$ the trained VAE (orange squares), and $(iii)$ the independent model (gray triangles), on a log-linear scale. The VAE reproduces the empirical $P(q)$ across the full support including the tail, whereas the independent model collapses to a narrow distribution. On the bottom right ({\bf f}), we plot the generalized Hopfield energy functional given in Eq.~\eqref{eq:generalized-hopfield} with its first two moments removed to visualize non-quadratic contributions. Clearly both patterns have significant higher-order contributions in the tails, indicating that the neural population has more complicated interactions than the usual two-body overlaps in the standard Hopfield model.}
    \label{fig:neural}
\end{figure*}

\subsection{Tensor order parameter: the Maier-Saupe model}
\label{sec:tensor}

The Maier-Saupe model of the nematic-isotropic phase transition in liquid crystals \cite{MaierSaupe1959,deGennesProst1993} is defined by the Hamiltonian
\begin{equation}\label{eq:H-MS}
    H = -\frac{J}{N}\sum_{i<j} P_2(\hat{\vb{x}}_i \cdot \hat{\vb{x}}_j) \;,
\end{equation}
where $\hat{\vb{x}}_i \in S^2$ are unit directors, in contrast to the discrete spins $x_i \in \{\pm 1\}$ of Sec.~\ref{sec:scalar}, and $P_2(y) = (3 y^2 - 1)/2$ is the second Legendre polynomial. A Hubbard-Stratonovich decoupling of the quadratic form now introduces a $3 \times 3$ symmetric auxiliary tensor $\vb{Z}$ (the nematic analog of the scalar $z$) and gives the factorized form
\begin{equation}\label{eq:pxz-MS}
    p(\hat{\vb{x}}_i \mid \vb{Z}) \propto \exp\!\left(\tfrac{3\beta J}{2}\, \hat{\vb{x}}_i^T \vb{Z} \hat{\vb{x}}_i\right) \;,
\end{equation}
which is a Bingham distribution on the sphere \cite{Bingham1974,Kent1982}. Because $\vb{Z}$ is a $3 \times 3$ symmetric traceless matrix it has $5$ independent components, exactly the $5$ degrees of freedom of the physical nematic order tensor $\bar{\vb{Q}} = \langle \hat{\vb{x}} \hat{\vb{x}}^T\rangle - \vb{I}/3$.

\vspace{0.2cm}

We trained a VAE on $N = 64$ equilibrium Maier-Saupe samples at temperatures $T/J \in [0.05, 1]$ with a $5$-dimensional latent and a decoder outputting the Bingham parameters of $p(\hat{\vb{x}} | \vb{Z})$ (see Methods~\ref{sec:meth:VAE} for the implementation of the Bingham distribution). The VAE-generated scalar nematic order parameter $S(T)$ correctly reproduces the location of the first-order transition at $T_{NI}/J \approx 0.22$ (C2), as well as the susceptibility and the Binder cumulant. Moreover, we train a small MLP to recover the auxiliary tensor $\vb{Z}$ from the encoder's latent variable and we find that they are related with a coefficient of determination above $0.9$ in every latent dimension (Fig.~\ref{fig:vector-tensor}b), confirming the mean-field equivalence for tensor-valued order parameters as well (C3). Furthermore, the VAE still has a vanishingly small ${\rm TC}_{|\vb{z}}$ as can be seen in Fig.~\ref{fig:hierarchy}l. 

\section{Experimental Application: latent structure of retinal codes}
\label{sec:retina}

Retinal population responses can be captured quantitatively by maximum-entropy models that are mathematically equivalent to Ising models with pairwise interactions \cite{Schneidman2006}, and analogous constructions extend to larger populations when global activity constraints are included \cite{Tkacik2014}. Starting from the salamander retina recordings of Ref.~\cite{Schneidman2006}, we assemble $145$ repeated presentations of a $26.5\,\mathrm{s}$ natural-movie clip for the same $N=40$ ganglion cells (see Methods~\ref{sec:meth:retina} for preprocessing). Correlations in this preparation are concentrated near a $20\,\mathrm{ms}$ time scale \cite{Schneidman2006,Tkacik2014}, so we partition time into consecutive non-overlapping $20\,\mathrm{ms}$ bins; in each bin the population state is $\{x_1, x_2, \cdots, x_{40}\}$ with $x_i = 1$ if neuron $i$ fired and $x_i = -1$ otherwise by analogy with our previous spin-systems. Treating successive bins as independent samples, we train a VAE with a linear decoder and with two latent dimensions to reproduce these $40$-bit words.

\vspace{0.2cm}

The data cannot be described by a fully factorized independent model such as the one given in Eq.~(\ref{eq:indp-mean-field}), since population synchrony and overlaps are poorly described by this modeling \cite{Schneidman2006} as can be seen in Fig.~\ref{fig:neural}(d-e). Refs.~\cite{Schneidman2006,Tkacik2014} instead fit Ising-like models. These Ising-like models rely on fitting $\mathcal{O}(N^2)$ couplings describing the microscopic interactions between all pairs of neurons. Instead, we show that a two-dimensional latent with a conditionally independent decoder can match the empirical count distribution. Only $2$ variables suffice to correctly capture the correlations between all $N = 40$ neurons instead of the $\sim 800$ couplings learned in Ref.~\cite{Schneidman2006}. Furthermore, by choosing the decoder to be linear, we ensure that the model learned by the VAE is equivalent to a generalized Hopfield model \cite{krotov2016dense,Demircigil2017,Ramsauer2021} whose Hamiltonian is given by
\begin{equation} \label{eq:generalized-hopfield}
    H = - \vb{h} \cdot \vb{\sigma} - F\left(\sum_{i = 1}^N \xi^{\mu}_i \sigma_i\right) \;,
\end{equation}
where $F(\vb{u}) = \log \left\langle e^{\sum_\mu z_\mu u_\mu} \right\rangle$ is the cumulant generating function of a tilted latent distribution $\tilde{p}(z)$ (see Methods~\ref{sec:meth:linear-decoder} for a derivation). 

\vspace{0.2cm}

We generate synthetic words from the learned Hopfield model and compare summary statistics to the data. We check the frequency of a random 10-bit word in the data compared to the generated distribution and find excellent agreement, see Fig.~\ref{fig:neural}(d). Given that Ref.~\cite{Schneidman2006} successfully fitted a disordered Ising-like model to these data, we adopt the overlap $q_{\alpha \beta}$ as an order parameter for such a system
\begin{equation} \label{eq:overlap}
    q_{\alpha \beta} = \frac{1}{N} \sum_{i = 1}^{N} x^{(\alpha)}_i x^{(\beta)}_{i} \;,
\end{equation}
evaluated between two independently sampled configurations $\vb{x}^{(\alpha)}$ and $\vb{x}^{(\beta)}$  \cite{MPV1987}. We draw $50\,000$ samples from (i) the VAE, (ii) a held-out testing dataset, and (iii) the independent model, and estimate $P(q)$ for each (Fig.~\ref{fig:neural}e). The behavior of $P(q)$ gives us two important observations. First, $P(q)$ acts like an order parameter for a spin-glass system, so the VAE reproduction of the empirical $P(q)$ is direct evidence that this population admits a finite-size mean-field description (C2). Second, we see that the overlap decays more slowly than the random independent model, meaning the neural encodings are redundant, proof of some underlying structure. 

\vspace{0.2cm}

We now infer the meaning of the two order parameters that the model has learned. Since we used a linear decoder, i.e. $p_\theta({\vb{x}} | \vb{z}) \propto \exp[\vb{x} \cdot (\vb{W} \vb{z} + \vb{b})]$ we can directly read-off the learned patterns $\xi_1$ and $\xi_2$ from the rows of $\vb{W}$ and the external field $\vb{h}$ from $\vb{b}$, as shown in Fig.~\ref{fig:neural}(a-b-c). We can also compute numerically the cumulant generating function of $\tilde{p}_\psi(\vb{z})$ and compare it to a Gaussian baseline as shown in Fig.~\ref{fig:neural}(f). We can see that the interaction does not seem to be quadratic, so the regular Hopfield model would be a poor description of this neural dataset. Instead it seems like the neural dataset behaves roughly quadratic in the bulk but clearly has higher-order moments contributing in the tails. From the results in Ref.~\cite{krotov2016dense} this would imply that this neural population has a larger storage capacity than a standard Hopfield model.

\section{Discussion}
\label{sec:discussion}

We have established an equivalence between VAEs and finite-size mean-field approximations from statistical mechanics, sharpened by the capacity criterion of Eq.~(\ref{eq:capacity}). This identification has three consequences: (C1) low-dimensional regular VAEs cannot faithfully represent data with no mean-field description. For example, the two-dimensional Ising model. (C2) Success on a complex empirical distribution constitutes evidence for a finite-size mean-field description, as we have demonstrated on Curie-Weiss (Sec.~\ref{sec:scalar}), Hopfield (Sec.~\ref{sec:vector}), Maier-Saupe (Sec.~\ref{sec:tensor}), and retinal population codewords (Sec.~\ref{sec:retina}). (C3) When such a description exists, its underlying microscopic parameters can be read off directly from the trained decoder, which we exemplified by recovering the full Hopfield pattern matrix from the decoder's logit-space Jacobian and fitting the parameters of a generalized Hopfield model to neural population recordings.

\section{Acknowledgments}

We thank Doruk Efe Gökmen, and Megi Dervishi for critical feedback on the manuscript. This research was partly supported from the National Science Foundation through
the Physics Frontier Center for Living Systems (PHY2317138) as well as NSF (DMS-2235451) and Simons
Foundation (MPS-NITMB-00005320) grants to the NSF-Simons
National Institute for Theory and Mathematics in Biology (NITMB). V.V is a Chan Zuckerberg
Biohub Chicago Investigator. V.V. acknowledges partial support from the Army Research Office under grant W911NF-22-2-0109 and W911NF-23-1-0212.  This work was completed
in part with resources provided by the University of
Chicago’s Research Computing Center. V.V
acknowledge partial support from the France Chicago
center through a FACCTS grant. M. B acknowledges support
through Schmidt Sciences, LLC.

\bibliographystyle{apsrev4-2}
\bibliography{references}

\cleardoublepage

\section{Methods}
\label{sec:methods}

\subsection{VAE architecture} \label{sec:meth:VAE}

\paragraph*{Encoder (CNN).} The temperature is concatenated as an extra channel to every single spin. Then the network consists of two convolutional blocks with hidden channels $[32, 64]$, kernel size $3$, stride $2$, padding $1$. Each block applies $\mathrm{Conv}_{1,2{\rm d}} \to \mathrm{BatchNorm} \to \mathrm{LeakyReLU}(0.01)$. The flattened activations passed to two fully connected heads producing the encoder mean and log-variance. The log-variance is clamped to $[-20, 20]$ for numerical stability.

\paragraph*{Spline prior.} The prior $p_\psi(\vb{z})$ is the pushforward of $\mathcal{N}(0, \mathbb{I}_d)$ through $K = 4$ stacked rational-quadratic neural spline layers \cite{Durkan2019} with $8$ knots and identity tails outside the bound $B = 5.0$. The spline parameters at each layer are produced by a per-layer hypernetwork $\mathrm{Linear}(1, 32) \to \mathrm{GELU} \to \mathrm{Linear}(32, n_{\rm params})$ conditioned on the temperature, with the output layer initialised to zero so that the prior coincides with $\mathcal{N}(0, \mathbb{I}_d)$ at the start of training. Because the spline prior is non-Gaussian, the KL divergence in the ELBO is computed by the change-of-variables formula on the reparameterised latent rather than in closed form. Due to our specific architectural choice the learned prior remains always factorizable, i.e. 
\begin{equation}
    p_\psi(\vb{z}) = \prod_{\mu = 1}^P p_\psi^{(\mu)}(z_\mu) \;.
\end{equation}

\paragraph*{Decoder (mirror CNN).} A linear map $\mathrm{Linear}(d, \, 64 \cdot N_{\rm pooled}^{\rm dim})$ reshapes onto the pooled spatial grid, followed by $\mathrm{ConvTranspose}(64 \to 32) \to \mathrm{ConvTranspose}(32 \to 32) \to \mathrm{Conv}(32 \to C_{\rm out})$ with kernel $3$, stride $2$, padding $1$, and output padding $1$. 

\paragraph*{Decoder (linear).} Only for the neural application we purposefully restrict the decoder to a single linear layer with a bias in order to make the VAE distribution formally equivalent to a generalized Hopfield model. 

\paragraph*{Likelihood and loss.} For Ising, Curie-Weiss, Hopfield and the neural application the decoder outputs $C_{\rm out} = 2$ logits per site over $\{-1, +1\}$, and the likelihood is the per-site cross-entropy summed over sites. For Maier-Saupe the decoder outputs $C_{\rm out} = 8$ Bingham parameters per site (a six-degree-of-freedom rotation $R \in SO(3)$ via Gram-Schmidt and two cumulative-softplus eigenvalues $\lambda_1 \leq \lambda_2 \leq 0$), and the per-site negative log-likelihood is evaluated by one-dimensional Gauss-Legendre quadrature. The total loss is
\begin{equation} \label{eq:loss-hp}
    \mathcal{L} \;=\; \sum_{i = 1}^{N} \mathrm{CE}_i \;+\; \beta \cdot D_{\rm KL}\big(q_\phi(\vb{z} \mid \vb{x}) \,\|\, p_\psi(\vb{z})\big) \,,
\end{equation}
or the Bingham analogue, with $\beta = 1$ throughout (no warm-up).

\subsection{Training and ELBO} \label{sec:meth:elbo}

The optimisation schedule is shared across all systems: AdamW with learning rate $10^{-3}$ and weight decay $10^{-5}$; an exponential learning-rate scheduler with $\gamma = 0.99$ per epoch; mixed-precision \texttt{bf16} arithmetic via PyTorch Lightning, with no gradient clipping. Training runs on a single GPU for a maximum of $100$ epochs with the model checkpoint selected on validation loss. The global random seed is $42$. The validation split is $10\%$ of the dataset, taken at the level of Zarr chunks (the last $10\%$ of chunks per temperature), so that no Monte-Carlo trajectory bleeds across the train/validation boundary. Per-system architectural and dataloader hyperparameters (latent dimension $d$, input channels $C_{\rm in}$, spatial size $L$, batch size, decoder type and output channels $C_{\rm out}$) are tabulated in Table~\ref{tab:architecture}.

\begin{table*}[t]
\centering
\caption{Per-system VAE architecture and dataloader hyperparameters. All systems share the same spline-flow encoder (Conv blocks $[32, 64]$, kernel $3$, stride $2$, padding $1$; $K = 4$ rational-quadratic spline layers with $8$ knots and bound $B = 5.0$), the same optimiser (AdamW, lr $10^{-3}$, weight decay $10^{-5}$, exponential schedule $\gamma = 0.99$), and the same training budget ($100$ epochs, seed $42$, bf16 mixed-precision). Temperature conditioning is on for the four physical systems and off for the retina (where there is a single condition).}
\label{tab:architecture}
\begin{tabular}{lccccccc}
\toprule
System & $d$ & $C_{\rm in}$ & $L$ & spatial dim & batch & decoder & $C_{\rm out}$ \\
\midrule
2D Ising         & $1$   & $1$ & $8$  & $2$ & $1024$ & CNN & $2$ \\
Curie-Weiss      & $1$   & $1$ & $64$ & $1$ & $1024$ & CNN & $2$ \\
Hopfield         & $P$\textsuperscript{$\ast$} & $1$ & $64$ & $1$ & $1024$ & CNN & $2$ \\
Maier-Saupe      & $5$   & $3$ & $64$ & $1$ & $2048$ & Bingham CNN  & $8$ \\
Retina           & $2$   & $1$ & $40$ & $1$ & $1024$ & Linear       & $2$ \\
\bottomrule
\end{tabular}
\\[4pt]
{\footnotesize\textsuperscript{$\ast$}\,$P \in \{1, 2, 4, 8, 16\}$ trained; the main-text results use $P = 4$.}
\end{table*}

\subsection{Equilibrium sampling} \label{sec:meth:sampling}

Training sets are drawn from the equilibrium Gibbs measure of the corresponding Hamiltonian. Two-dimensional Ising samples are produced by a Numba-accelerated Wolff cluster sampler \cite{Wolff1989}, Curie-Weiss samples by direct draws from the analytic single-site marginal, Hopfield and Maier-Saupe sample using the Hubbard-Stratonovich block Gibbs sampler described below. Common settings across systems are random seed $42$, coupling $J = 1$, and a measurement interval of $10$ Monte-Carlo sweeps. Per-system temperature grids, chain counts, and equilibration sweeps are tabulated in Table~\ref{tab:sampling}.

\begin{table*}[t]
\centering
\caption{Per-system Monte-Carlo sampling parameters. The 2D Ising and Hopfield samplers exploit the global $\mathbb{Z}_2$ symmetry of the Hamiltonian by storing each measured configuration together with its negation, doubling the on-disk sample count; the same total of $\sim 10^5$ configurations per temperature is collected in all four systems. The Curie-Weiss model is sampled directly from the closed-form magnetisation distribution $P(M) \propto \binom{N}{(N+M)/2}\exp(\beta J M^2 / 2N)$, so no MCMC chain is needed.}
\label{tab:sampling}
\begin{tabular}{lcccccccc}
\toprule
System & sampler & $T$ range & spacing & $\#T$ & $L$ & chains & $n_{\rm burn}$ & $n_{\rm meas}$ \\
\midrule
2D Ising      & Wolff (Numba)              & $[0.1, 3.5]$  & log    & $50$ & $8$  & $1$    & $5\!\times\!10^5$ & $5\!\times\!10^4$ \\
Curie-Weiss   & direct (exact)             & $[0.1, 3.5]$  & linear & $50$ & $64$ & ---    & ---               & $10^{5}$              \\
Hopfield      & HS block Gibbs             & $[0.1, 3.5]$  & log    & $50$ & $64$ & $10^3$ & $2\!\times\!10^3$ & $5\!\times\!10^4$ \\
Maier-Saupe   & HS block Gibbs             & $[0.05, 1.0]$ & log    & $50$ & $64$ & $10^3$ & $2\!\times\!10^3$ & $10^5$            \\
\bottomrule
\end{tabular}
\end{table*}

\paragraph*{Hubbard-Stratonovich Gibbs sampler for the Hopfield and Maier-Saupe models.} Single-spin Metropolis on the bare Hamiltonian $H_{\rm Hopfield}$ of Eq.~(\ref{eq:H-hopfield}) crosses between competing pattern wells through configurations of vanishingly small equilibrium weight near the retrieval transition, which inflates the autocorrelation time of the order parameter. We instead sample in the augmented space $(\vb{x}, \vb{z})$, where the auxiliary Hubbard-Stratonovich field $\vb{z}$ provides direct access to the pattern overlaps and makes those crossings cheap. Applying the Hubbard-Stratonovich identity to Eq.~(\ref{eq:H-hopfield}) gives the augmented joint
\begin{equation} \label{eq:HS-joint}
    p(\vb{x}, \vb{z}) \;\propto\; \exp\!\left( -\tfrac{\beta J N}{2} |\vb{z}|^2 + \beta J \sum_i (\vb{\xi} \vb{z})_i \, x_i \right) \,,
\end{equation}
with the two block-Gibbs conditionals
\begin{align}
    p(\vb{z} \mid \vb{x}) &= \mathcal{N}\!\left( \vb{m}(\vb{x}),\; (\beta J N)^{-1} \mathbb{I}_P \right) \,, \\
    m_\mu(\vb{x}) &= \frac{1}{N} \sum_i \xi_i^\mu x_i \,, \label{eq:gibbs-z} \\
    p(x_i = +1 \mid \vb{z}) &= \frac{\exp\!\left( \beta J \,(\vb{\xi} \vb{z})_i \right)}{2 \cosh (\beta J (\xi \vb{z})_i)} \,. \label{eq:gibbs-x}
\end{align}
The sites $x_i$ are conditionally independent given $\vb{z}$, so all $N$ updates in Eq.~(\ref{eq:gibbs-x}) factorize and are evaluated in one parallel sweep. Each iteration costs $\mathcal{O}(N P)$, dominated by the matrix-vector product $\vb{\xi} \vb{z}$, two orders of magnitude cheaper than a comparable single-spin Metropolis sweep with rejection. For each temperature we initialize $10^{3}$ parallel chains each with $\vb{x}^{(0)}$ at a random Gaussian configuration, discard a burn-in of $2 \cdot 10^3$ sweeps, and retain one configuration every $10$ sweeps thereafter until $5 \times 10^4$ samples are collected.

\vspace{0.2cm}

An identical construction can be made for the Maier-Saupe model by alternating between drawing an auxiliary $3\times3$ tensor from a Gaussian distribution and then per-site Bingham sampling with directors given by the auxiliary tensor. 

\subsection{Pattern recovery via logit-space Jacobian} \label{sec:meth:recovery}

At the variational optimum the trained Hopfield decoder reproduces the conditional of Eq.~(\ref{eq:pxz-hopfield}) up to an invertible reparametrization $R \in GL(P)$ of the latent space, so that its log-odds at site $n$ read
\begin{equation} \label{eq:methods-logit}
    f_\theta(\vb{z})_n \;\equiv\; \log\frac{p_\theta(x_n = +1 \mid \vb{z})}{p_\theta(x_n = -1 \mid \vb{z})} \;=\; 2 \beta J \,(\vb{\xi} R^{-1} \vb{z})_n \,.
\end{equation}
Differentiating Eq.~(\ref{eq:methods-logit}) with respect to $z_\mu$ gives the logit-space Jacobian
\begin{equation} \label{eq:methods-jacobian}
    \chi_{n \mu}(\vb{z}) \;=\; \pdv{f_\theta(\vb{z})_n}{z_\mu} \;=\; 2 \beta J \,(\vb{\xi} R^{-1})_{n \mu} \,,
\end{equation}
which is independent of $\vb{z}$.

\vspace{0.2cm}

To unwind the gauge $R$ we exploit the fact that the rows of $\vb{\xi}$ are i.i.d.\ Rademacher vectors, i.e.\ maximally non-Gaussian. We draw $512$ samples $\vb{z}^{(k)} \sim p_\psi(\vb{z} \mid T)$ from the trained conditional prior at the working temperature, evaluate the per-sample Jacobian using automatic differentiation of the decoder, and feed the averaged matrix $\bar{J} = \langle J(\vb{z}) \rangle_{p_\psi}$ to FastICA \cite{Hyvarinen1999} with unit-variance whitening and $P$ components. The $P$ resulting sources $\hat{\vb{s}}^\nu \in \mathbb{R}^N$ are real-valued, non-Gaussian directions in site space; we set $\hat{\xi}_n^{\nu} = \mathrm{sign}(\hat{s}_n^{\nu})$ to recover binary patterns, retaining the best of five FastICA restarts in mean absolute overlap with the trial reference $\vb{\xi}$. ICA returns the patterns only up to the $\mathbb{Z}_2^P \rtimes S_P$ symmetries (column sign and column permutation) under which the Hopfield Hamiltonian of Eq.~(\ref{eq:H-hopfield}) is invariant by construction. We resolve this residual ambiguity by computing the overlap matrix $O_{\mu \nu} =  \hat{\vb{\xi}}^{\nu} \cdot \vb{\xi}^{\mu}  / N$ and solving the linear assignment $\sigma^\star = \arg\max_{\sigma \in S_P} \sum_\mu |O_{\mu\, \sigma(\mu)}|$ by the Hungarian algorithm \cite{Kuhn1955}, followed by a per-pattern sign flip aligning $\hat{\vb{\xi}}^{\sigma^\star(\mu)}$ with $\vb{\xi}^\mu$. The aligned overlap matrix is reported in Fig.~\ref{fig:vector-tensor}(a) for $P = 4$, $N = 64$ at $T/J = 0.5$. 

\subsection{Total correlation metric} \label{sec:meth:TC}
The total correlation was estimated following the method detailed in Ref.~\cite{Xu2025}, which we summarize here for completeness. The method stems from the variational re-parametrization of the Kullback-Leibler divergence \cite{Nguyen2010} which states
\begin{equation}
    D_{\rm KL}(P || Q) = \sup_f \left\{ \mathbb{E}_P[f({\vb X})] - \mathbb{E}_Q[e^{f({\vb X}) - 1}] \right\} \;,
\end{equation} 
which allows us to recover the total correlation as defined in Eq.~(\ref{eq:TC_KL}) by learning a function $f_\theta(X)$ using a neural network optimized to minimize the loss:
\begin{equation} \label{eq:f-divergence-loss}
    \mathcal{L}_{\rm TC} = 1 + \sum_{i = 1}^{K_1} \log f_{\vb \theta}({\bf X_i}) - \sum_{j = 1}^{K_2} f_{\vb \theta}({\bf Y_j}) \;,
\end{equation}
where $\{{\bf X_i}\}$ are samples from the joint $P({\vb X}) = p(\vb{x} \mid z)$ and $\{{\vb Y_j}\}$ are samples from the mixture $\tfrac{1}{2}[P({\vb Y}) + Q({\vb Y})]$, with $Q({\vb Y}) = \prod_i p(y_i \mid z)$ the independent reference. The unique global minimizer of Eq.~(\ref{eq:f-divergence-loss}) is then the bounded \emph{relative} density ratio of Eq.~(\ref{eq:RDR}),
\begin{equation} \label{eq:RDR-methods}
    r(x_1, \cdots, x_N) = \frac{P({\vb{x}})}{\tfrac{1}{2}\!\left[ P({\vb{x}}) + Q({\vb{x}}) \right]} \;\in\; [0, 2] \;,
\end{equation}
which is the choice advocated by Ref.~\cite{Xu2025} as a numerically stable surrogate for the unbounded standard ratio $P/Q$. The learned machine learning model $f_\theta$ will approximate this relative ratio, $f_\theta \simeq r$; refer to Ref.~\cite{Xu2025} for details on convergence guarantees. Provided we have trained $f_\theta$ successfully, we can then compute the total correlation by simply evaluating
\begin{equation} \label{eq:TC-estimator}
    {\rm TC}_{|z} \;=\; \left\langle \log \frac{r(\vb{X})}{2 - r(\vb{X})} \right\rangle_{\vb{X} \sim P} \;,
\end{equation}
where the rewriting $\log(P/Q) = \log[r/(2-r)]$ follows directly from Eq.~(\ref{eq:RDR-methods}) and the empirical average is taken over the same samples $\{\vb{X_i}\} \sim P(\vb{x} \mid z)$ used in Eq.~(\ref{eq:f-divergence-loss}). The conditional total correlation reported in the right column of Fig.~\ref{fig:hierarchy} is obtained by further averaging Eq.~(\ref{eq:TC-estimator}) over $z \sim p_\psi(\vb{z} \mid T)$ at each temperature.

\vspace{0.2cm}

Since we have access to the full density ratio $r(\vb{x})$ and not only integrated metrics such as $D_{\rm KL}$, we can fit a linear regression to check which relevant observables ($\langle x_i \rangle$, $\langle x_i x_j \rangle$, \ldots) contribute most to the mismatch between $P$ and $Q$. The resulting residual covariance matrices are shown in the insets of Fig.~\ref{fig:hierarchy}: the VAE trained on the two-dimensional Ising model reproduces all one-point observables but fails on two-point and higher correlators, whereas Curie-Weiss, Hopfield and Maier-Saupe residuals are featureless.

\subsection{Retinal preprocessing} \label{sec:meth:retina}

The salamander retinal recordings of Ref.~\cite{Schneidman2006} consist of $N = 40$ ganglion cells stimulated by the same $26.5$\,s natural-movie clip on $145$ distinct trials. Spike sorting and unit selection are inherited from the original publication; we make no further unit rejection.

\paragraph*{Time binning.} Spike trains are binned into disjoint windows of width $\Delta t = 20$\,ms, giving $T = \lfloor 26.5\,\mathrm{s} / 20\,\mathrm{ms} \rfloor \sim 1325$ bins per trial. The bin width matches the dominant time scale of the population autocorrelation \cite{Schneidman2006,Tkacik2014}, beyond which successive bins decorrelate. In each bin $t$ of trial $\tau$ we form a binary word $\vb{x}^{(t, \tau)} \in \{-1, 1\}^{40}$ with $x_i^{(t, \tau)} = 1$ if neuron $i$ fired at least once in the bin, else $x_i^{(t, \tau)} = -1$. Multi-spike events within a single bin are rare ($< 2\%$ of nonzero entries at the population mean firing rate) and contribute the same value as single-spike events; the binarisation therefore preserves the leading population statistics by construction. The total dataset contains $145 \times 1325 \approx 1.92 \times 10^5$ words.

\paragraph*{Train/test split.} We hold out 10\% of the bins chosen uniformly at random as a test set and use the remaining samples for training. The test set therefore contains $\approx 10^4$ words. The overlap distribution $P(q)$ of Fig.~\ref{fig:neural}(e) is evaluated by drawing $50\,000$ independent pairs of words from the test set, and the same number of independent pairs from VAE samples and from the matched-marginals independent model.

\paragraph*{Sample independence.} The VAE is trained on the assumption that successive bins are i.i.d.\ samples of the steady-state population code. This is an idealisation: across the $20$\,ms boundary residual temporal correlations remain, both intrinsic to the retinal circuit and inherited from the smooth natural stimulus. 

\subsection{Equivalence between linear decoder VAE and generalized Hopfield} \label{sec:meth:linear-decoder}

For the retinal application we purposefully restrict the decoder to be linear, in such a case the jpdf of the VAE is given by
\begin{equation} \label{eq:jpdf-linear-decoder}
    p(\vb{x}) = \int \dd \vb{z} \; p_\psi(\vb{z}) \prod_{i = 1}^N \frac{\exp\left(x_i \sum_{\mu} W_{i \mu} z_\mu + x_i b_i\right)}{2 \cosh(\sum_{\mu} W_{i \mu} z_\mu + b_i)} \;.
\end{equation}
We can manipulate Eq.~\eqref{eq:jpdf-linear-decoder} to re-write it as
\begin{equation}
    p(\vb{x}) = e^{\vb{b} \cdot \vb{x}} \int \dd \vb{z} \; \tilde{p}(\vb{z}) \exp[ \vb{x} \cdot \sum_{\mu} \vb{W_\mu} z_\mu] \;,
\end{equation}
where we introduced the tilted distribution
\begin{equation}
    \tilde{p}(\vb{z}) = \frac{p_\psi(\vb{z})}{ \prod_{i = 1}^N 2 \cosh(\sum_\mu W_{i\mu} z_\mu + b_i) } \;.
\end{equation}
Then the Hamiltonian $H$ is
\begin{equation} \label{eq:generalized-hopfield-2}
    H = - \log p(\vb{x}) = - F(\vb{W} \cdot \vb{x}) - \vb{b} \cdot \vb{x} \;,
\end{equation} 
where $F(\vb{u})$ is the cumulant generating function of the distribution $\tilde{p}(\vb{z})$. The Hamiltonian in Eq.~\eqref{eq:generalized-hopfield-2} closely recovers the definition provided in Ref.~\cite{krotov2016dense} with the added generalization that we have an overal biasing term $\vb{b} \cdot \vb{x}$ and the function $F(\vb{u})$ could in theory contain cross-coupling terms. 

\subsection{Capacity criterion} \label{sec:meth:capacity}

For any bipartition $\{1, \dots, N\} = A \sqcup B$ of the sites, the conditional independence ansatz of Eq.~(\ref{eq:cond-indp}) implies that the sub-configurations $\vb{X}_A$ and $\vb{X}_B$ are conditionally independent given $\vb{Z}$, so $\vb{X}_A \to \vb{Z} \to \vb{X}_B$ forms a Markov chain. The data-processing inequality then gives 
\begin{equation} \label{eq:meth:dpi}
    I(\vb{X}_A ; \vb{X}_B) \;\leq\; I(\vb{X}_A ; \vb{Z}) \;\leq\; I(\vb{X} ; \vb{Z}) \,, 
\end{equation}
for any bipartition $A \sqcup B$. Maximizing the left-hand side over bipartitions yields our first bound,
\begin{equation} \label{eq:meth:rate-lower}
    I(\vb{X} ; \vb{Z}) \;\geq\; I_{\rm bip}(p) \;\equiv\; \max_{A \sqcup B} I(\vb{X}_A ; \vb{X}_B) \,.
\end{equation}
Equation~(\ref{eq:meth:rate-lower}) is the \emph{message-passing bound}: all correlations between $A$ and $B$ must be routed through the single hub $\vb{Z}$, so the rate $I(\vb{X}; \vb{Z})$ of the latent channel is at least the maximum information that must cross any cut. The quantity $I_{\rm bip}(p)$ is a property of the target distribution alone, so Eq.~(\ref{eq:meth:rate-lower}) translates a physical fact about $p(\vb{x})$, namely the scaling of its correlations, into a requirement on the network.

\vspace{0.2cm}

The rate that the network actually delivers is the KL term of the ELBO in Eq.~(\ref{eq:elbo}),
\begin{equation} \label{eq:meth:rate-kl}
    R \;\equiv\; \langle D_{\rm KL}(q_\phi(\vb{z} \mid \vb{x}) \,\|\, p_\psi(\vb{z})) \rangle_{p(\vb{x})} \,,
\end{equation}
which coincides with $I(\vb{X} ; \vb{Z})$ up to a non-negative discrepancy that vanishes at the ELBO optimum \cite{HoffmanJohnson2016,Alemi2018,Tishby2015}.

\vspace{0.2cm}

For a Gaussian prior $p_\psi(\vb{z}) = \mathcal{N}(0, \mathbb{I}_d)$ and Gaussian encoder $q_\phi(\vb{z} \mid \vb{x}) = \mathcal{N}(\vb{\mu}(\vb{x}), \sigma(\vb{x})^2 \mathbb{I}_d)$ in the near-deterministic regime $\sigma \ll 1$, the dominant term reduces to the \emph{sphere-packing capacity}
\begin{equation} \label{eq:meth:rate-sphere}
    R \;\simeq\; d \log(1/\sigma) \,,
\end{equation} 
which is the logarithm of the number $\sim (1/\sigma)^d$ of radius-$\sigma$ balls that fit into the unit prior; the latent channel can distinguish at most $e^R$ equivalence classes of inputs, and any correlation structure beyond this budget is lost into the encoder noise. 

\vspace{0.2cm}

Combining Eq.~(\ref{eq:meth:rate-lower}) and Eq.~(\ref{eq:meth:rate-sphere}) gives a \emph{necessary} condition for faithful VAE reconstruction,
\begin{equation} \label{eq:meth:capacity}
    d \log(1/\sigma) \;\gtrsim\; I_{\rm bip}(p) \;,
\end{equation}
which trades the physics of the data (the right-hand side, set by the scaling of correlations in $p(\vb{x})$) against the architecture of the network (the left-hand side, set by $d$ and the achievable encoder noise). 

\subsubsection{Going beyond the Gaussian regime}

The above mentioned capacity criterion is restricted to a Gaussian prior $p_\psi(\vb{z}) = \mathcal{N}(0, \mathbb{I}_d)$ with a Gaussian posterior $q_\phi(\vb{z} | \vb{x}) = \mathcal{N}(\mu(\vb{x}), \sigma(\vb{x})^2 \mathbb{I}_d)$ in the near-deterministic regime $\sigma \ll 1$. Here we aim to relax these conditions. 

\vspace{0.2cm}

Firstly, we adress the learned prior $p_\psi(\vb{z})$ which is generally non-Gaussian. Concretely, the architecture of the main text takes $p_\psi(\vb{z}) = g_\psi \# \mathcal{N}(0, \mathbb{I}_d)$, i.e. the pushforward of a unit Gaussian through a learned diffeomorphism $g_\psi : \mathbb{R}^d \to \mathbb{R}^d$. Because $g_\psi$ is a diffeomorphism, mutual information is invariant under it: $I(\vb{X}; \vb{Z}) = I(\vb{X}; \vb{U})$ with $\vb{U} = g_\psi^{-1}(\vb{Z}) \sim \mathcal{N}(0, \mathbb{I}_d)$. Every flow-prior VAE is therefore equivalent, as a joint distribution, to a Gaussian-prior VAE with redefined decoder $\tilde p_\theta(\vb{x} \mid \vb{u}) \equiv p_\theta(\vb{x} \mid g_\psi(\vb{u}))$, and the message-passing bound of Eq.~(\ref{eq:meth:rate-lower}) and hence the capacity criterion of Eq.~(\ref{eq:capacity}) carry over verbatim. 

\vspace{0.2cm}

The role of the flow is instead to remove a mismatch between the aggregated posterior $q_\phi(\vb{z}) \equiv \int \dd \vb{x} \, q_\phi(\vb{z} \mid \vb{x}) p(\vb{x})$ and the prior. Following the ELBO surgery of Refs.~\cite{HoffmanJohnson2016,Alemi2018}, the rate of Eq.~(\ref{eq:meth:rate-kl}) decomposes as
\begin{equation} \label{eq:hoffman-johnson}
    R \;=\; I(\vb{X}; \vb{Z}) \;+\; D_{\rm KL}(q_\phi(\vb{z}) \,\|\, p_\psi(\vb{z})) \,,
\end{equation}
where the first term is the useful rate that enters Eq.~(\ref{eq:meth:rate-lower}), and the second is an \emph{excess-KL tax} that the network pays whenever its aggregated posterior fails to match the prior. With a unit-Gaussian prior and a physically multimodal aggregated posterior, as in the low-temperature Curie-Weiss phase of Fig.~\ref{fig:vae_arch}(b) or the broken-symmetry Hopfield retrieval phase, the second term is strictly positive and the encoder is forced into a regularisation compromise. A sufficiently expressive flow drives $D_{\rm KL}(q_\phi(\vb{z}) \,\|\, p_\psi(\vb{z})) \to 0$ and thereby returns the full budget $R$ to useful rate.

\vspace{0.2cm}

Secondly, we adress the near-deterministic regime restriction. The full expression for $R$ given in Eq.~\eqref{eq:rate-kl} can be computed exactly when both distributions are gaussian and is given by
\begin{equation} \label{eq:R}
    R = \frac{1}{2} \langle ||\mu(\vb{x})||^2 \rangle + \frac{d}{2} [ \langle \sigma(\vb{x})^2 \rangle - 1 - \langle \log \sigma(\vb{x})^2 \rangle ] \;.
\end{equation}
Notice indeed that when $\sigma \ll 1$ the dominant term comes from $\sim d \log(1 / \sigma)$. Instead when the dominant contribution comes from the first term in Eq.~\eqref{eq:R} would correspond to the well-known {\it variance collapse} of VAEs, i.e. each input-point is uniquely mapped to far-away points in the latent space. The condition in Eq.~\eqref{eq:capacity} can be more generally stated using Eq.~\eqref{eq:R} and Eq.~\eqref{eq:hoffman-johnson}
\begin{equation}
    R - D_{\rm KL}(q_\phi(\vb{z}) \,\|\, p_\psi(\vb{z})) \gtrsim I_{\rm bip}(p) \;.
\end{equation}
We can now see why pushing $\mu \to +\infty$ in Eq.~\eqref{eq:R} does not salvage expressivity. All the different collapse modes of VAE networks will typically correspond to maximizing $R$ at the cost of blowing up $D_{\rm KL}(q_\phi(\vb{z}) \,\|\, p_\psi(\vb{z}))$. Hence, most practical regimes (i.e. non-collapsed VAEs and regular ELBO loss) are well described by the limit in Eq.~\eqref{eq:capacity}. 

\subsection{Information-theoretic capacity bound} \label{sec:meth:practical-capacity}

With Eq.~(\ref{eq:capacity}) in hand, the question of which equilibrium distributions are amenable to a VAE reduces to estimating how $I_{\rm bip}(p)$ scales with system size. For the mean-field models of the main text the Hubbard-Stratonovich representation provides an explicit CI decomposition through a $d$-dimensional auxiliary field, which we use to upper-bound $I_{\rm bip}$ via the data-processing chain of Eq.~(\ref{eq:meth:rate-lower}); for the two-dimensional Ising model no such decoupling exists, and we lower-bound $I_{\rm bip}$ instead by the classical bipartite area law.

\vspace{0.2cm}

\paragraph*{Hubbard-Stratonovich rate bound.} Each mean-field model considered admits a representation $p(\vb{x}) = \int \dd \vb{z} \, p(\vb{z}) \prod_i p(x_i \mid \vb{z})$ with an exponential-family conditional that is linear in $\vb{z}$ and a Gaussian prior centred on the mean-field saddle. Bayes' rule in this representation gives, to leading order in $N$,
\begin{equation} \label{eq:HS-posterior}
    p(\vb{z} \mid \vb{x}) \;=\; \mathcal{N}\!\left( \vb{m}(\vb{x}), \, (\beta J N)^{-1} \mathbb{I}_d \right) \,,
\end{equation}
where $\vb{m}(\vb{x})$ is the $d$-dimensional order parameter read off from the linear coupling (magnetisation for Curie-Weiss, overlap vector for Hopfield, nematic tensor for Maier-Saupe), and the posterior width $\sim (\beta J N)^{-1/2}$ is set by the Gaussian fluctuations of the HS action around the saddle. Subtracting the posterior entropy from the prior entropy and combining with Eq.~(\ref{eq:meth:rate-lower}) yields
\begin{equation} \label{eq:Ibip-MF-bound}
    I_{\rm bip}(p_{\rm MF}) \;\leq\; \tfrac{d}{2} \log N + \mathcal{O}(1) \,,
\end{equation}
which is comfortably satisfied by a VAE with a $d$-dimensional latent and predicts that the variance of the latent should scale as $\sim 1/\sqrt{N}$. Note also that as expected in the thermodynamic limit, i.e. $N \to +\infty$, the system effectively becomes deterministic with $\vb{z}$ having a vanishing variance.

\vspace{0.2cm}

\paragraph*{Two-dimensional Ising at criticality.} The local Ising Hamiltonian admits no finite-dimensional decoupling auxiliary field, and Eq.~(\ref{eq:Ibip-MF-bound}) does not apply. Instead, the classical bipartite mutual information between two halves of an $L \times L$ box at the Onsager transition obeys a lower bound of area-law form,
\begin{equation} \label{eq:Ibip-Ising}
    I_{\rm bip}^{\rm Ising}(T_c) \;\gtrsim\; c\, L + \mathcal{O}(\log L) \,, \qquad c = \mathcal{O}(1) \,,
\end{equation}
established numerically and analytically in Refs.~\cite{Wilms2011,Iaconis2013,Stephan2014}. Eq.~(\ref{eq:capacity}) then becomes $d \log(1/\sigma) \gtrsim L \sim \sqrt{N}$, which implies $\sigma < \exp(-\sqrt{N}/d)$ and at $N = 64$ as in the main text already approaches the machine-precision limit in fp16. Realistically any training with latent dimension $d$ that does not scale with $\sqrt{N}$ will be incredibly unstable and most likely will not converge: the VAE fails on the two-dimensional Ising model near criticality not through a poor choice of hyperparameters, but because the physical correlations of $p(\vb{x})$ transport more information than the low-dimensional latent channel can carry.

\end{document}